\documentclass[11pt]{article}
\usepackage{graphicx}

\begin{document}

\def\xslash#1{{\rlap{$#1$}/}}
\def \p {\partial}
\def \dd {\psi_{u\bar dg}}
\def \ddp {\psi_{u\bar dgg}}
\def \pq {\psi_{u\bar d\bar uu}}
\def \jpsi {J/\psi}
\def \psip {\psi^\prime}
\def \to {\rightarrow}
\def\bfsig{\mbox{\boldmath$\sigma$}}
\def\DT{\mbox{\boldmath$\Delta_T $}}
\def\xit{\mbox{\boldmath$\xi_\perp $}}
\def \jpsi {J/\psi}
\def\bfej{\mbox{\boldmath$\varepsilon$}}
\def \t {\tilde}
\def\epn {\varepsilon}
\def \up {\uparrow}
\def \dn {\downarrow}
\def \da {\dagger}
\def \pn3 {\phi_{u\bar d g}}

\def \p4n {\phi_{u\bar d gg}}

\def \bx {\bar x}
\def \by {\bar y}

\begin{center}
{\Large\bf  Single Spin Asymmetry in Lepton Angular Distribution of Drell-Yan Processes }
\par\vskip20pt
J.P. Ma$^{1,2}$ and G.P. Zhang$^{3}$     \\
{\small {\it
$^1$ Institute of Theoretical Physics, Academia Sinica,
P.O. Box 2735,
Beijing 100190, China\\
$^2$ Center for High-Energy Physics, Peking University, Beijing 100871, China  \\
$^3$ School of Physics,  Peking University, Beijing 100871, China
}} \\
\end{center}
\vskip 1cm
\begin{abstract}
We study the single spin asymmetry in the lepton angular distribution of Drell-Yan processes in the framework of collinear factorization.
The asymmetry has been
studied in the past and different results have been obtained. In our study we take an approach different than that
used in the existing study.
We explicitly calculate the transverse-spin dependent part of the differential cross-section
with suitable parton states. Because the spin is transverse, one has to take multi-parton states
for the purpose. Our result agrees with one of the existing results.
A possible reason for the disagreement with others is discussed.
\vskip 5mm
\noindent
% PACS numbers:
\end{abstract}
\vskip 1cm
\par
Single Spin Asymmetry(SSA) can in general exist in high energy scattering with a transversely polarized hadron. The existence of such an asymmetry implies the existence of helicity-flip interactions and nonzero
absorptive part of scattering amplitudes. Therefore,
the experimental and theoretical study of SSA  offers a new way to explore
the inner-structure of hadrons. Because of its importance, significant effort has been devoted to the
study of SSA. Reviews about this research field can be found in \cite{Review}.
\par
Theoretical predictions for SSA can be made by using the concept of QCD factorization, if large
momentum transfers exist in a process. In the framework of QCD collinear factorization, the nonperturbative
effect of the transversely polarized hadron is factorized into matrix elements of twist-3 operators, as pointed out
in \cite{QiuSt,EFTE}. In this work we will focus on the SSA in the lepton angular distribution of Drell-Yan processes.
SSA in this case has been studied with collinear factorization in \cite{DY1,DY2,DY3,DY4,ZM,AT},
but different results have been obtained. The purpose of our work is to solve the discrepancy with
a method which is different than that employed in the past studies.
\par
We consider the Drell-Yan process
\begin{equation}
  h_A ( P_A, s) + h_B(P_B) \to \gamma^* (q) +X \to  \ell^-(k_1)  + \ell ^+ (k_2) + X,
\end{equation}
where $h_A$ is a spin-1/2 hadron with the spin-vector $s$. $h_B$ is unpolarized.
We take a light-cone coordinate
system in which $h_A$ moves in the $+$-direction and $h_B$ moves in the $-$-direction.
$h_A$ is transversely polarized with the spin vector $s^\mu =(0,0,\vec s_\perp)$. We employ the Collins-Soper
frame to describe the lepton angular distribution\cite{CS-frame}. In this frame the lepton pair is in rest.
We take the spin direction as the direction of the $x$-axis and denote the solid angle of $\ell^-$
in the frame as $\Omega$. We denote the invariant mass of the lepton pair as $Q$ and
$S=(P_A+P_B)^2$. We define the following SSA  relative to the spin direction:
\begin{equation}
A_N = \left (\frac{d\sigma( \vec s_\perp) }{d Q^2 d\Omega} - \frac{d\sigma( -\vec s_\perp) }{d Q^2 d\Omega} \right )
   \biggr /
 \left ( \frac{d\sigma( \vec s_\perp) }{d Q^2 d\Omega} + \frac{d\sigma( -\vec s_\perp) }{d Q^2 d\Omega} \right ).
\label{AN}
\end{equation}
As mentioned, the existing results for this asymmetry are different. From \cite{DY2,DY3} the result reads:
\begin{equation}
A_N^{\cite{DY2,DY3}}
    = -\frac{\sin(2\theta)\sin\phi } {Q(1+\cos^2\theta)}
    \frac{ \displaystyle{  \int d x d y T_F(x,x) \bar q(y) \delta ( xyS-Q^2) } }
      {  \displaystyle { \int dx dy q(x) \bar q(y) \delta ( xyS-Q^2) } }.
\label{BQ}
\end{equation}
In the above $T_F(x,x)$ is a twist-3 matrix element of $h_A$, whose definition will be given later.
$\bar q(y)$ is the antiquark parton distribution of $h_B$. In \cite{DY1} the derived $A_N$ has a derivative
term of $T_F$ in addition to the above expression. Later, the asymmetry has been re-studied in \cite{ZM, AT}.
From \cite{ZM} $A_N$ is only the half of that given in Eq.(\ref{BQ}), while the study in \cite{AT} confirms
the result from \cite{DY2,DY3}.
\par
The result for the defined asymmetry has a number of interesting aspects in comparison with the asymmetry
defined with other differential cross section, which has been studied extensively in \cite{JQVY1,KKnew,KY3G}.
The asymmetry studied in \cite{JQVY1,KKnew,KY3G} is at the order of $\alpha_s$ and has different contributions.
In contrast, the asymmetry in Eq.(\ref{AN}) is at the order of $\alpha_s^0$ and is predicted in a simple form.
Because of these it likely provides the best way to access twist-3 matrix elements by measuring $A_N$.
Therefore, it is important to solve the theoretical discrepancy of this asymmetry. It should be noted that
all mentioned results of $A_N$ are derived with the method of diagram expansion at hadron level, in which one works
directly with hadron states to evaluate differential cross-sections of hadrons.
\par
It should be realized that QCD factorizations are general properties of QCD, if they are proven.
In principle one can derive the factorization for a hadron scattering by replacing hadrons with QCD partons.
After the replacement one can explicitly calculate
the differential cross section of the corresponding parton scattering and relevant matrix elements of QCD operators.
With the obtained results, one can directly derive
perturbative coefficient function in the factorization, hence the factorization.
It has been started in \cite{MS1,MS2,MS3,MS4,MS5} with this approach to derive the factorization of SSA.
In this work we will take this approach to study the collinear factorization of $A_N$.
\par
Before we turn to our calculation of SSA with partonic states, we give the definition of the twist-3 matrix element
appearing in $A_N$. We will use the  light-cone coordinate system, in which a
vector $a^\mu$ is expressed as $a^\mu = (a^+, a^-, \vec a_\perp) =
((a^0+a^3)/\sqrt{2}, (a^0-a^3)/\sqrt{2}, a^1, a^2)$ and $a_\perp^2
=(a^1)^2+(a^2)^2$. In the system we introduce two light-cone vectors: $n^\mu =(0,1,0,0)$ and $l^\mu =(1,0,0,0)$.
Other notations are:
\begin{equation}
  g_\perp^{\mu\nu} = g^{\mu\nu} - n^\mu l^\nu - n^\nu l^\mu,
  \ \ \ \ \ \
  \epsilon_\perp^{\mu\nu} =\epsilon^{\alpha\beta\mu\nu}l_\alpha n_\beta, \ \ \ \
  \epsilon^{0123}=1.
\end{equation}
The definition of the twist-3 matrix element $T_F(x_1,x_2)$ reads:
\begin{eqnarray}
T_F (x_1,x_2)
   =    -g_s \tilde s_\mu \int \frac{dy_1 dy_2}{4\pi}
   e^{ -iy_2 (x_2-x_1) P^+ -i y_1 x_1 P^+ }
   \langle P_A, \vec s_\perp \vert
           \bar\psi (y_1n ) \gamma^+ G^{+\mu}(y_2n) \psi(0) \vert P_A,\vec s_\perp \rangle
\label{tw3}
\end{eqnarray}
with $\tilde s^\mu = \epsilon^{\mu\nu}_\perp s_{\perp\nu}$. The above definition is given
in the light-cone gauge $n\cdot G=0$. In other gauge gauge links along the direction $n$
should be supplemented to make the definition gauge invariant. Instead of the spin-vector
one can also use helicity $\lambda$
to describe the polarization of $h_A$. In this case the general forward scattering amplitude like
$\langle h_A(\lambda')\vert {\mathcal O }\vert h_A(\lambda)\rangle$ with some operator ${\mathcal O}$
is a $2\times 2$ matrix in the helicity space. It is clear that the non-diagonal part corresponds to
the forward scattering amplitude of $h_A$ with the transverse polarization.
\par
\begin{figure}[hbt]
\begin{center}
\includegraphics[width=4cm]{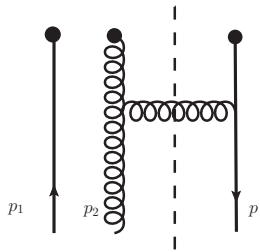}
\end{center}
\caption{The diagram for $T_F (x,x)$ at one-loop. The black dots denote the insertion of operators
used to define $T_F(x_1,x_2)$ in Eq.(\ref{tw3}). }
\label{SGP-Tq}
\end{figure}
\par
\par
If we use a transversely polarized single-quark state to replace $h_A$ in the definition of $T_F$,
one will always get $T_F=0$ because the helicity of a massless quark is conserved in QCD. In order
to study SSA and its factorization, one has to consider multi-parton states for the replacement\cite{MS3}.
Following \cite{MS3} we consider the state
\begin{equation}
 \vert n [\lambda ] \rangle  =  \vert q(p,\lambda)  \rangle + c_1
                   \vert q(p_1,\lambda_q) g(p_2,\lambda_g ) [\lambda =\lambda_q +\lambda_g ] \rangle +\cdots,
\label{MPS}
\end{equation}
with $p_1+p_2 =p$. The state $\vert n [\lambda ] \rangle$ is a superposition of a single quark- and quark-gluon state.
It has the helicity $\lambda$. The single quark state must have the same helicity $\lambda$, while in the quark-gluon state, denoted
as $qg$-state, the sum of the quark helicity $\lambda_q$ and the gluon helicity $\lambda_g$ must be $\lambda =\lambda_q +\lambda_g$.  The $qg$-state
is in the fundamental representation of $SU(N_c)$-gauge group. We take the momenta as $p_1=x_0 p$ and $p_2=(1-x_0) p$
with $p^\mu =P_A^\mu =(P_A^+, 0,0,0)$. If we replace $h_A(\lambda)$ in Eq.(\ref{tw3}) with the state
$\vert n [\lambda ] \rangle$, one will find that $T_F$ receives contributions only from the matrix elements
of the interference between the single quark- and the $qg$-state, i.e., $\langle q(\lambda_q)\vert {\mathcal O }
\vert q(\lambda_q)g(\lambda_g)\rangle$ or $\langle q(\lambda_q)g(\lambda_g) \vert {\mathcal O }
\vert q(\lambda_q)\rangle$. It is noted that the total helicity in the bra- and ket-state is different,
but the quark always has the same helicity. In general, one can use the state in Eq.(\ref{MPS})
to construct a $2 \times 2$ spin density matrix $\langle n[\lambda'] \vert {\mathcal O}\vert n[\lambda]\rangle$
in helicity space for a given operator ${\mathcal O}$,
as discussed in detail in \cite{MS3,MS4,MS5}. By taking corresponding operators, e.g., the one used
to define $T_F$, one can obtain from the non-diagonal part the transverse-spin dependent matrix element
with ${\mathcal O}$.
\par
With the state $\vert n [\lambda ] \rangle$ instead of $h_A$ in Eq.(\ref{tw3}) one can calculate $T_F$
perturbatively. The tree-level result can be found in \cite{MS3,MS4}.
At this order the matrix element $T_F(x,x)$ relevant in this work is zero.
$T_F(x,x)$ becomes nonzero at one-loop.
As found in \cite{MS3,MS4}, at one-loop level
there is only one diagram giving nonzero contribution to $T_F (x,x)$ in the light-cone-
or Feynman gauge.
The calculation of the diagram is straightforward. The contribution has an U.V.- and a collinear divergence. Both
are regularized with the dimensional regularization as poles of $\epsilon=4-d$. After extracting
the U.V. pole  we have\cite{MS3,MS4}:
\begin{equation}
T_F (x,x, \mu ) =- \frac{ g_s \alpha_s}{4} N_c (N_c^2-1) x_0 \sqrt{2x_0} \delta (x_0-x)
 \left [ \left (-\frac{2}{\epsilon_c} \right ) + \ln\frac{e^\gamma \mu^2}{4\pi \mu_c^2} \right ]
 + {\mathcal O}(g_s\alpha_s^2) ,
\label{TF}
\end{equation}
where the pole is the collinear divergence with the index $c$. $\mu$ is the renormaliation scale related
to the U.V. pole, and $\mu_c$ is that related to the collinear pole. For simplicity we have taken $c_1=1$
in Eq.(\ref{MPS}).
\par
Now we turn to the Drell-Yan process in Eq.(1). The relevant hadronic tensor is defined as:
\begin{equation}
W^{\mu\nu}  = \sum_X \int \frac{d^4 x}{(2\pi)^4} e^{iq \cdot x} \langle h_A (P_A, s_\perp), h_B(P_B)  \vert
    \bar q(0) \gamma^\nu q(0) \vert X\rangle \langle X \vert \bar q(x) \gamma^\mu q(x) \vert
     h_B(P_B),h_A (P_A, s_\perp)  \rangle,
\label{WT}
\end{equation}
where $q$ is the momentum of the lepton pair.
The differential cross section appearing in Eq.(\ref{AN}) is related to the tensor as:
\begin{equation}
\frac{d\sigma}{d Q^2 d\Omega} = \frac{\alpha^2}{S  Q^4} \int d^4  q \delta (q^2-Q^2)
\left [k_1^\mu k_2^\nu + k_1^\nu k_2^\mu - k_1\cdot k_2 g^{\mu\nu} \right ] W_{\mu\nu}.
\label{dsigma}
\end{equation}
It should be noted that in the defined distribution in Eq.(9) only the invariant mass $Q$ of the lepton
pair is fixed and some components of $q$ are integrated, e.g., $q_\perp$. In the integration one should note that
the lepton momenta $k_{1,2}$ depend on the momentum $q$ in the moving frame and on the solid angle $\Omega$ in Collins-Soper frame. Because the integration over $q$ with $q^2$ fixed, the integration can give some soft divergences in the small $q_\perp$ region.
Therefore one should perform the factorization of the defined differential cross-section in Eq.(\ref{dsigma}) instead of  structure functions of $W^{\mu\nu}$. Only in the case with other observables which are directly related to structure functions, one needs to
perform factorizations for these functions.
\par
We denote the spin-dependent and symmetric part of $W^{\mu\nu}$ as
$\hat W^{\mu\nu}$. Only this part will give contributions to $A_N$.
For $\vec q_\perp\neq 0$ the tensor can be decomposed into eight structure functions
\cite{PR,AMS}. These structure functions are in general singular with $\vec q_\perp \to 0$.
In principle one can have a part of $\hat W^{\mu\nu}$ which is proportional to $\delta^2(\vec q_\perp)$. This
part has a simple form:
\begin{equation}
\hat W^{\mu\nu} = \left [ \tilde s^\mu \left ( \frac{P_A^\nu}{P_A\cdot q}
   - \frac{P_B^\nu}{P_B\cdot q} \right ) + \tilde s^\nu \left ( \frac{P_A^\mu}{P_A\cdot q}
   - \frac{P_B^\mu}{P_B\cdot q}\right ) \right ] \hat W_0 + \cdots,
\end{equation}
with $\hat W_0 \propto \delta^2(\vec q_\perp)$. The $\cdots$ represent the terms of the
eight structure functions which can be found in \cite{PR,AMS}.
One expects that these structure functions can have a factorized form. In \cite{JQVY1,KKnew,KY3G},
the trace part, i.e., $\hat W^\mu_{\ \mu}$ has been studied with collinear factorization.
In \cite{MS3,MS4,MS5}, the factorization of the same part has been studied with multi-parton states.
It should be noted that the factorization of structure functions can be different than that of $A_N$.
The structure functions are for fixed $\vec q_\perp$. For $A_N$, i.e., for the differential cross
section in Eq.(\ref{dsigma}) $\vec q_\perp$ is integrated over. The collinear- and I.R. divergences
in Eq.(\ref{dsigma}) can appear in a different way than those in structure functions. This will be
clearly seen in the results obtained with the multi-parton state.
\par
Having the result for $T_F(x,x)$ of the state $\vert n(p,\lambda)\rangle$ in Eq.(\ref{TF}) , we need to calculate the asymmetry
of the scattering with the same state $\vert n(p,\lambda)\rangle$ to study the factorization or to check the factorization
in Eq.(\ref{BQ}). For this we consider the scattering:
\begin{equation}
    n(p,\lambda ) + \bar q(\bar p) \to \gamma^*(q) + X,
\label{PP}
\end{equation}
with $P_A =p$ and $P_B=\bar p$.
The leading order of SSA here is at $g_s\alpha_s$. As we will see, the obtained differential cross-section contains collinear divergences.
We will show that the collinear divergences can be factorized with the collinear divergence in $T_F(x,x)$
of Eq.(\ref{TF}).
Because  $T_F(x,x)$ is at order of $g_s\alpha_s$, it results in that the perturbative coefficient function
is at order of $\alpha_s^0$ as in Eq.(\ref{BQ}).
The finite contributions in the  differential cross-section can be factorized with the tree-level result
of $T_F(x_1,x_2)$ which is at order of $g_s$. Hence, the finite contributions
will delivery corrections at order of $\alpha_s$ to the perturbative coefficient function.
Hence, we only need to find collinear divergences in the differential cross-section at the leading but nontrivial order.
\par
\par
%%%%%%%%%%%%%%%% Inset Fig. 1 here %%%%%%%%%%%%%%%%%%%%%%%%%%%%%%
\begin{figure}[hbt]
\begin{center}
\includegraphics[width=12cm]{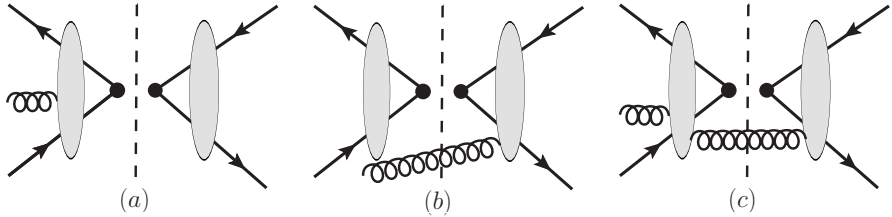}
\end{center}
\caption{Classes of diagrams for the contributions to the hadronic tensor with the multi-parton state in Eq.(\ref{MPS}). The black dots are insertion of electromagnetic current operators.
The broken line is the cut. }
\label{Types}
\end{figure}
\par
The contributions to $\hat W^{\mu\nu}$ can be classified into three classes of Feynman diagrams for $\hat W^{\mu\nu}$ ,
which are given in Fig.2. In each diagram of Fig.2 there is a cut dividing the diagram into a left-
and right part. In order to have SSA, a cut should be also exist in the left- or right part. This cut is not drawn in Fig.2.
In Class (a) of diagrams, represented with Fig.2a, there is no parton in the intermediate
state. In Class (b) of diagrams represented with Fig.2b, the initial gluon without interactions with
partons in the left part of the diagram, goes through the cut to interact partons in the right part.
The intermediate state only contains this gluon. It is clear that contributions from these two classes of diagrams
are proportional to $\delta^2(\vec q_\perp)$. Fig.2c represnts diagrams of Class (c). In these diagrams
the intermediate state contains an emitted gluon. Hence, in contributions from Class (c) $\vec q_\perp$ can be nonzero.
It is possible to have an additional class of contributions, which are those diagrams where the initial gluon in the left part in Fig.2c can go without interactions
through the cut. But at the leading order, there is no absorptive part in the left- or right part. Hence the contributions from this additional class are zero at the order.

\par
Not all classes of diagrams need to be considered. One can show that the contributions from Class (b) are exactly zero.
For Fig.2b we denote the left- and right part as the amplitude ${\mathcal T}^{\mu}_L$ and ${\mathcal T}^{\nu}_R$,
respectively. These amplitudes are in fact the matrix elements:
\begin{eqnarray}
{\mathcal T}^\mu_L  =\langle \bar 0 \vert J^\mu \vert q(p_1,\lambda_q) \bar q(\bar p) \rangle, \ \ \
{\mathcal T}^\nu_R  =\langle q(p,\lambda_q ) \bar q(\bar p) \vert J^\nu \vert  g(p_2,\lambda_g)\rangle,
\end{eqnarray}
where we have labeled the helicity of the quark and gluon explicitly. It should be noted that
The $q\bar q$ in ${\mathcal T}^\mu_L$ is in color-octet, i.e., the pair carries the same color as that of the gluon
in ${\mathcal T}^\nu_R$.
These amplitudes can be decomposed into the form factors:
\begin{eqnarray}
{\mathcal T}^\mu_L  = \bar v(\bar p) \gamma^{\mu} T^a  u(p,\lambda_q ) F_1(q^2),
\ \ \ \
{\mathcal T}^\nu_R  = \left ( \frac{p_1^\nu}{q\cdot p_1} -\frac{\bar p^\nu}{q\cdot \bar p} \right )
    \bar u (p,\lambda_q) T^a  \gamma\cdot \epsilon(\lambda_g ) v(\bar p) G_1 (q^2) ,
\label{FF}
\end{eqnarray}
where the color index $a$ is the color of the gluon, $\epsilon(\lambda_g)$ is the polarization vector
of the gluon and $q$ is fixed as $q=p_1+\bar p$. The form factors
$F_1$ and $G_{1}$ are complex functions of $q^2$ in general. Using these expressions one can calculate $\hat W^{\mu\nu}$
directly. One easily finds:
\begin{equation}
  \hat W^{\mu\nu}_{(b)} \propto \left ( 1 +\lambda_q \lambda_g \right ).
\end{equation}
Since we take the state in Eq.(\ref{MPS}) as a spin-1/2 system because $h_A$ is with spin-1/2,
one always has $\lambda_q \lambda_g =-1$. Therefore, the contributions from Class (b) are zero.
This holds at any order of $\alpha_s$. In fact, the conclusion about the contributions from Class (b)
is a consequence of the helicity conservation of QCD.
\par
\par
%%%%%%%%%%%%%%%% Inset Fig. 2 here %%%%%%%%%%%%%%%%%%%%%%%%%%%%%%
\begin{figure}[hbt]
\begin{center}
\includegraphics[width=12cm]{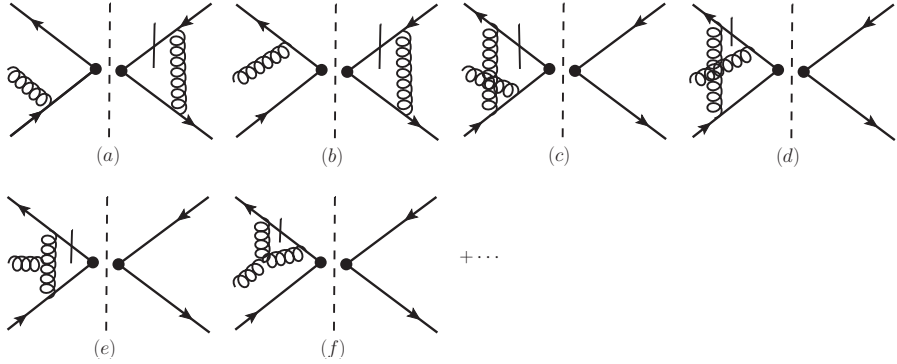}
\end{center}
\caption{The diagrams of Class(a) contributions. These diagrams contains soft divergences. The diagrams
which do not contain soft divergences are represented with $\cdots$. }
\label{ClassA}
\end{figure}
\par
There are many diagrams for contributions from Class (a). Since we are interested in soft divergences, we only need to consider those diagrams
which contain soft divergences. Those diagrams are given in Fig.3. In each diagram, a short bar cutting a quark propagator means to take the absorptive part of the propagator. In fact the short bar implies one or more physical cuts in
the amplitude of the left- or right part of the diagrams. E.g., when we extend the short bar to the bottom of Fig. 2a, the right
part represents in fact the scattering $\gamma^* \to q\bar q \to q\bar q$, hence a nonzero absorptive part of the amplitude
is generated. A special care should be taken for the left part of Fig.3a, because two collinear partons merge into a quark propagator. This brings up an ambiguity like $0/0$, when two partons are exactly collinear. To deal the ambiguity we first take the gluon momentum $p_2$ off-shell by giving it a small $-$-component, i.e., $p_2^\mu = (p_2^+, p_2^-,0,0)$.
Then the left part of Fig.3a becomes:
\begin{equation}
{\mathcal T}^{\mu}_{(3a,L)} = g_s \bar v(\bar p) \gamma^\mu \frac{\gamma\cdot (p_1+p_2)}{(p_1+p_2)^2+i\varepsilon} T^a \gamma\cdot \epsilon T^a u(p_1)=g_s \bar v(\bar p) \gamma^\mu \frac{\gamma^+ p_2^- }{ 2p_2^- (p_1+p_2)^+ +i\varepsilon} T^a \gamma\cdot \epsilon T^a u(p_1),
\end{equation}
from the above we can in the last step take $p_2^- =0$. This is equivalent to take the quark propagator as
the special quark propagator given in \cite{tw4}.
\par
The soft divergences in Fig.3 can easily be worked out. Here we notice that the soft divergences in these diagrams
are not collinear divergences. They are generated through exchange of a Glauber gluon, whose momentum $k$ has the pattern
$k^\mu \sim (\lambda^2,\lambda^2,\lambda,\lambda)$ with $\lambda \to 0$. The reason for this momentum pattern
is the physical cuts in amplitudes. With the on-shell conditions from the cuts the $+$- and $-$-component
of $k$ must be at order of $\lambda^2$ if $k_\perp$ is at order of $\lambda$ with $\lambda\to 0$.
For the hadronic tensor we also need to add
the contributions of conjugated diagrams of Fig.3. We have the divergent part regularized with dimensional regularization:
\begin{eqnarray}
\tilde W^{\mu\nu}\biggr\vert_{6a+6b} &=&  g_s \alpha_s \frac{(N_c^2-1)^2}{4 N_c^2} \delta^4 (p+\bar p-q)
      \sqrt{2 x_0} \tilde s^{\{\nu}  \left ( \bar p -p \right )^{\mu \}} \left ( -\frac{2}{\epsilon_G} \right ) + \cdots,
\nonumber\\
\tilde W^{\mu\nu}\biggr\vert_{6e+6f} &=& -g_s \alpha_s \frac{N_c^2-1}{4} \delta^4 (p+\bar p-q)
\sqrt{2 x_0}  \tilde s^{ \{\nu}  \left ( \bar p  -  p \right ) ^{\mu\}} \left ( -\frac{2}{\epsilon_G} \right )  + \cdots ,
\nonumber\\
\tilde W^{\mu\nu}\biggr\vert_{6c+6d} &=&  g_s \alpha_s \frac{N_c^2-1}{4 N_c^2} \delta^4 (p+\bar p-q)
\sqrt{2 x_0}  \tilde s^{ \{\nu}  \left ( \bar p - p \right ) ^{\mu\}} \left ( -\frac{2}{\epsilon_G} \right )  + \cdots ,
\label{Class-A}
\end{eqnarray}
with $\epsilon_G =4-d$. Its pole represents the Glauber divergence. The terms represented
by $\cdots$ are finite. We have used the notation to denote the symmetric part of a tensor built by two vectors:
\begin{equation}
  A^{\{\mu} B^{\nu\}} = A^\mu B^\nu + A^\nu B^\mu.
\end{equation}
From Eq.(\ref{Class-A}) the sum of the three pieces of contributions are finite. Therefore, we conclude
that the contributions of Class (a) to the hadronic tensor and hence to the differential cross-section
are finite at the leading order.
\par
\par
%%%%%%%%%%%%%%%% Inset Fig. 2 here %%%%%%%%%%%%%%%%%%%%%%%%%%%%%%
\begin{figure}[hbt]
\begin{center}
\includegraphics[width=11cm]{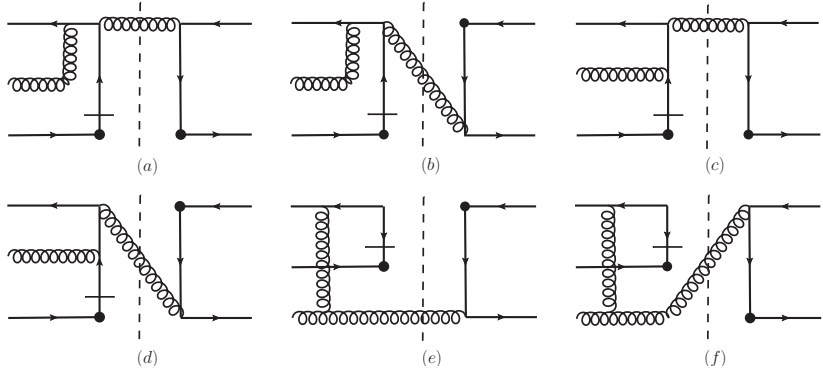}
\end{center}
\caption{The diagrams for Class (c). }
\label{Feynman-dg4}
\end{figure}
\par
Now we turn to the contributions from Class (c). At the leading order the contributions come from diagrams given in Fig.4.
These contributions to $\hat W^{\mu\nu}$ are nonzero and finite for $\vec q_\perp\neq 0$.  But when we calculate
the differential cross-section, a divergence appears after the integration over $q_\perp$. To find the divergence
we can in the first step expand those contributions from Fig.4 in $q_\perp \sim \lambda$ with $\lambda\to 0$ and keep only those
contributions which will give the divergence after the integration over $q_\perp$. From Eq.(\ref{dsigma})
the contributions to $\hat W^{\mu\nu}$ at order of $\lambda^{-1}$ or higher will give finite contributions
to the differential cross-section.
\par
By expanding the contribution from each diagram in Fig.4, we find
that all contributions except those from Fig.4e are at order of $\lambda^{-1}$ or higher order. Therefore, we have
the divergent part of $\hat W^{\mu\nu}$ of Class (c):
\begin{eqnarray}
\tilde W^{\mu\nu}\biggr\vert_c  &=& -\frac{ g_s \alpha_s}{4 \pi} (N_c^2-1) \sqrt{2 x_0}
    \delta (1-y)\delta (x-x_0) \biggr [
       \frac{1}{(q_\perp^2)^2}  x_0 \tilde s \cdot q_\perp g_\perp^{\mu\nu}+ \frac{1}{2 p\cdot \bar p  q^2_\perp} x_0 p^{\{\mu} \tilde s^{\nu\}}
\nonumber\\
      && -\frac{1}{(q_\perp^2)^2 p\cdot \bar p } \left (  \tilde s \cdot q_\perp \bar p^{\{\mu} q^{\nu\}}_\perp
 + \frac{1}{2} q^2_\perp \bar p^{\{\mu}\tilde s^{\nu\}} \right )\biggr ]
             + {\mathcal O}(\lambda^{-1}).
\label{F4E}
\end{eqnarray}
In the above,  the contribution with $g^{\mu\nu}_\perp$ is at order of $\lambda^{-3}$ and
the contribution from the remaining three terms  is at order of $\lambda^{-2}$. Only the terms in the first line
of Eq.(\ref{F4E}) give divergent contributions
to the differential distribution in Eq.(9) when we integrate over $q_\perp$.
The terms in the second line do not give divergent contributions because of the rotation covariance.
The terms at $\lambda^{-1}$ will only give finite contributions.
\par
Adding all contributions together we have $\hat W^{\mu\nu}$ at the leading order:
\begin{eqnarray}
\tilde W^{\mu\nu}  &=& -\frac{ g_s \alpha_s}{4 \pi} (N_c^2-1) \sqrt{2 x_0} \biggr \{  (n-l)^{\{\mu}\tilde s^{\nu\}}
   \delta (1-x)\delta (1-y) \delta^2(\vec q_\perp ) \hat w_0
\nonumber\\
 && + \delta (x-x_0) \delta (1-y) \biggr [
       \frac{1}{(q_\perp^2)^2}  x_0 \tilde s \cdot q_\perp g_\perp^{\mu\nu}+ \frac{1}{2 p\cdot \bar p  q^2_\perp} x_0 p^{\{\mu} \tilde s^{\nu\}}
\nonumber\\
  &&  -\frac{1}{(q_\perp^2)^2 p\cdot \bar p } \left (  \tilde s \cdot q_\perp \bar p^{\{\mu} q^{\nu\}}_\perp
 + \frac{1}{2} q^2_\perp \bar p^{\{\mu}\tilde s^{\nu\}} \right )\biggr ]
      + {\mathcal O}(\lambda^{-1}) \biggr \} + {\mathcal O}(g_s\alpha_s^2) .
\label{WT}
\end{eqnarray}
$\hat w_0$ is a finite constant.
Our result of is $U(1)$-electromagnetic gauge invariant
up to order of $\lambda^{-1} $. This can be checked by noting $q^\mu = p_1^\mu + \bar p^\mu +q^\mu_\perp
+ {\mathcal O}(\lambda^2)$:
\begin{equation}
  q_\mu \tilde W^{\mu\nu} = q_\nu \tilde W^{\mu\nu} = 0 +
  {\mathcal O}(\lambda^{-1}).
\end{equation}
We notice that the spin-independent part of the hadronic tensor $W^{\mu\nu}$ is proportional to $\delta^2(\vec q_\perp)$
at the leading order of $\alpha_s$, because only the process $q\bar q \to \gamma^* $
contributes. This is in contrast to the case studied in Eq.(\ref{WT}), where one has a contribution with $\vec q_\perp \neq 0$.

\par
Substituting the result in Eq.(\ref{WT}) into Eq.(\ref{dsigma}) and performing the integration over $q_\perp$,
we find
\begin{eqnarray}
\frac{d\sigma( \vec s_\perp) }{d Q^2 d\Omega} =
\left  [ \frac{ g_s \alpha_s}{4} (N_c^2-1) x_0 \sqrt{2 x_0}\left (-\frac{2}{\epsilon_c} \right ) \right ]  \frac{\delta(x_0 s-Q^2)}{128 \pi^2 Q^3}
 \sin(2\theta)\sin\phi
  +\cdots,
\label{ddsig}
\end{eqnarray}
where the divergence is regularized with $\epsilon_c=4-d$ and $\cdots$ denote finite contributions. The finite contributions are from the
term with $\hat w_0$ and those
at order ${\mathcal O}(\lambda^{-1})$ in Eq.(\ref{WT}). We notice here
that the upper limit of $q_\perp$ in the integration  is finite by the energy-momentum conservation.
In Eq.(\ref{ddsig})
we take the spin-direction in the $x$-direction, and the moving direction of $\ell^-$ in the Collins-Soper frame
is given by $(\sin\theta\cos\phi, \sin\theta\sin\phi,\cos\theta)$.

\par
The divergence comes from the $q_\perp$-region with $q_\perp\sim \lambda \to 0$. In this region one easily finds that
the gluon exchanged at the bottom of Fig.4e is collinear to the incoming gluon. Its momentum $k$
scales like $k^\mu \sim {\mathcal O}(1,\lambda^2,\lambda,\lambda)$ with $\lambda\to 0$. Therefore, the divergence
is a collinear one. With the result of $T_F(x,x)$ given in Eq.(\ref{TF}) and the leading result for an antiquark distribution
in an antiquark, i.e., $\bar q(y) =\delta (1-y) +{\mathcal O}(\alpha_s)$, we can derive the factorized result for the differential cross-section:
\begin{eqnarray}
\frac{d\sigma( \vec s_\perp) }{d Q^2 d\Omega} =-\frac{ \alpha^2 }{8 Q^3 N_c}
 \sin(2\theta)\sin\phi  \int d x d y T_F(x,x) \bar q(y) \delta ( xyS-Q^2),
\end{eqnarray}
and the asymmetry:
\begin{equation}
A_N
    = -\frac{\sin(2\theta)\sin\phi } {2 Q(1+\cos^2\theta)}
    \frac{ \displaystyle{  \int d x d y T_F(x,x) \bar q(y) \delta ( xyS-Q^2) } }
      {  \displaystyle { \int dx dy q(x) \bar q(y) \delta ( xyS-Q^2) } }.
\end{equation}
This is our main result. In our results we have taken the electric charge fraction $e_q$ of quark as $1$. One can easily generalize
the above results to any flavor. The finite contribution in Eq.(\ref{ddsig}) will be factorize with $T_F(x_1,x_2)$ at tree-level
and give a correction at order of $\alpha_s$ to the above $A_N$. Here, we have a case that from an partonic observable at a given order
of $\alpha_s$ the extracted perturbative coefficient functions can be at different orders.
This is the nontrivial order-mixing as observed in \cite{MS4}.
\par

Before we make a comparison with existing results, we show in the below that the differential cross-section
is indeed factorized with $T_F(x,x)$.
For this we write down explicitly the contribution of Fig.4e:
\begin{eqnarray}
{\mathcal M} \biggr\vert_{4e } &=&- \frac{g_s^3}{2N_c (2\pi)^4} f^{abc}
  \pi \delta ((p_1-q)^2) \delta (\bar p^- -q^- -k^-)  \frac{ \pi  }{  k^ + (p-k)^2(p_2-k)^2}
\nonumber\\
    && {\rm Tr} \left \{   \biggr [  T^a  u(p_1) \bar u (p) T^b \gamma_\rho \gamma\cdot (p-k) \biggr ] \gamma^\mu
                 v(\bar p) \bar v(\bar p)
         \gamma_{\rho_1} T^c \gamma\cdot (p_1-q) \gamma^\nu  \right \}
\nonumber\\
    && \left ( (p_2+k)^{\rho_1} \epsilon^\rho +(-2k +p_2)\cdot \epsilon g^{\rho\rho_1}
        +(-2p_2+k)^\rho \epsilon^{\rho_1} \right ) ,
\label{F4EA}
\end{eqnarray}
where $k$ is the momentum of the gluon crossing the cut with $k^+ = p^+ -q^+$ and $\vec k_\perp =-\vec q_\perp$.
We have performed the integration over $k^-$ with the on-shell condition of the gluon. This gives
$k^- =k^2_\perp/(2k^+)$. For finding the collinear divergence one can safely neglect $k^-$ in the $\delta$-function
for $-$-components of momenta.
The product in $[\cdots ]$ in the above is a matrix
with  color- and Dirac indices. This matrix can be expanded with $\gamma^\mu$and $\gamma^\mu \gamma_5$
for Dirac indices and with the color matrix $T^a$ for color indices. We can write the product as:
\begin{eqnarray}
 f^{abc} \biggr [  T^a  u(p_1) \bar u (p) T^b \gamma_\rho \gamma\cdot (p-k) \biggr ] & = & f^{abd}{\rm Tr}
 \left ( T^d T^a T^b \right )
   T^c \left ( \gamma^ -  A^{+}_{\ \rho}  + {\mathcal B}_\rho \right) ,
\nonumber\\
A^{+}_{\ \rho} &=& \bar u (p)  \gamma_\rho \gamma\cdot (p-k) \gamma^+ u(p_1).
\label{deco}
\end{eqnarray}
In the above we have explicitly written out the structure $\gamma^-$ for Dirac indices, the remaining terms have been
denoted as ${\mathcal B}_\rho$. If we only keep the term with $A^{+}_{\ \rho}$
and take $\gamma_{\rho_1}$ in Eq.(\ref{F4EA}) as $\gamma\cdot l n_{\rho_1}$,
 i.e.,
\begin{eqnarray}
{\mathcal M} \biggr\vert_{4e} &=&- \frac{1}{2N_c(2\pi)^4 } \pi \delta ((p_1-q)^2) \delta (\bar p^- -q^-) {\rm Tr}
  \left ( T^c  \gamma^- \gamma^\mu     v(\bar p) \bar v(\bar p)
         \gamma^- T^c \gamma\cdot (p_1-q) \gamma^\nu \right )
\nonumber\\
   &&         \biggr \{ \frac{ \pi  }{k^+ (p-k)^2(p_2-k)^2}
           A^{+}_{\ \rho}  n_{\rho_1}  f^{abc}{\rm Tr}
 \left ( T^c T^a T^b \right )
\nonumber\\
     && \left ( (p_2+k)^{\rho_1} \epsilon^\rho +(-2k +p_2)\cdot \epsilon g^{\rho\rho_1}
        +(-2p_2+k)^\rho \epsilon^{\rho_1} \right )  \biggr \}   + \cdots,
\label{fac}
\end{eqnarray}
where $\cdots$ represent the remaining contributions from ${\mathcal B}_\rho$ in Eq.(\ref{deco}) or $\gamma_{\rho_1}$ in Eq.(\ref{F4EA})
with $\rho_1 =\perp$.
We have:
\begin{eqnarray}
\tilde W^{\mu\nu}  &=& -\frac{ g_s \alpha_s}{4 \pi} (N_c^2-1) \sqrt{2 x_0}
    \delta (1-y)\delta (x-x_0) \biggr [
       \frac{1}{(q_\perp^2)^2}  x_0 \tilde s \cdot q_\perp g_\perp^{\mu\nu}
 + \frac{1}{2 p\cdot \bar p q^2_\perp}  x_0 p^{\{\mu} \tilde s^{\nu\}} \biggr ]
\nonumber\\
    &&  + {\mathcal O}(\lambda^{-2}) + {\mathcal O}(\lambda^{-1}).
\end{eqnarray}
The contributions in the first line come only from the term with $A^{+}_{\ \rho}$ in Eq.(\ref{deco}) and with $\gamma_{\rho_1}$ in Eq.(\ref{F4EA})  as  $\gamma\cdot l n_{\rho_1}$.
These contributions are exactly those in the first line of Eq.(\ref{F4E}), which generate the divergent contributions in the differential cross-section given in Eq.(\ref{ddsig}). The terms represented by ${\mathcal O}(\lambda^{-2})$ do not give divergent
contributions with the similar reason as discussed after Eq.(\ref{F4E}).
Now we note that the integral over $k_\perp$ with the part in $\{\cdots \}$ in Eq.(\ref{fac}) multiplied with
a factor $k^\mu_\perp$
is proportional the contribution to $T_F(x,x)$ given in Fig.1. The factor $k_\perp^\mu$ can come from the first line
of Eq.(\ref{fac}), or from the leptonic tensor in Eq.(\ref{dsigma}), when we express the lepton momenta
with the lepton momentum in the Collins-Soper frame and the momentum $q$. It should be noticed that
$\vec k_\perp$ in Eq.(\ref{fac}) is just $-\vec q_\perp$.
Therefore, the obtained divergence in the spin-dependent part of the differential cross-section
is exactly factorized with $T_F(x,x)$. With the above discussion, it is also clear that
for the differential cross-section in Eq.(\ref{dsigma}) one should perform the factorization only after the integration
over $q_\perp$.
\par
Our result disagrees with those in \cite{DY1,DY2,DY3,AT}. $A_N$ obtained here is only half of those derived in \cite{DY2,DY3,AT}.
Although we have used a method different than that in the existing studies, our result agrees with that given in \cite{ZM},
where the contributions factorized with chirality-odd operators are also given.
In \cite{AT}, the issue of gauge invariance in the problem has been emphasized. In our study,
gauge invariance is explicitly kept because we take on-shell parton states to calculate the hadronic tensor.
Our result is also $U(1)$-gauge invariant as checked in Eq.(20).
\par
We can explore the reason for the discrepancy.  In \cite{DY2,DY3,AT}
one factorizes the hadronic tensor in the first step. Then in the next step one uses the factorized tensor to calculate the differential cross-section by performing the
integration over $q_\perp$ indicated by Eq.(\ref{dsigma}), where the factorized tensor seems proportional
to $\delta^2(\vec q_\perp)$.
This procedure is only correct if the integration over $q_\perp$ does not generate soft divergences.
However, the integration does generate a collinear divergence as we have seen here.
This is the main reason for the discrepancy. To see the discrepancy more clearly, we re-arrange
our result in Eq.(\ref{WT}) as:
\begin{eqnarray}
\tilde W^{\mu\nu}  &=& -\frac{ g_s \alpha_s}{4 \pi} (N_c^2-1) \sqrt{2 x_0}
    \delta (1-y)\delta (x-x_0) \biggr [
 \frac{1}{2 p\cdot \bar p  q^2_\perp} \left ( x_0 p^{\{\mu} \tilde s^{\nu\}} -\bar p^{\{\mu}\tilde s^{\nu\}} \right )
\nonumber\\
  && + \frac{1}{(q_\perp^2)^2} \left ( x_0 \tilde s \cdot q_\perp g_\perp^{\mu\nu}
       -\frac{\tilde s \cdot q_\perp} {\bar p\cdot p}\bar p^{\{\mu} q^{\nu\}}_\perp \right ) \biggr ] + \cdots,
\end{eqnarray}
where $\cdots$ stand for terms which are irrelevant here or at higher order of $\alpha_s$.
 In \cite{DY2,DY3,AT} the factorized tensor used to calculate $A_N$ has only the tensor structure which is equivalent
 to the part given in the first line in the above. If we only use this part to calculate $A_N$, we obtain the same
 factorized result. It is noted here that the soft divergence of this part is proportional to $\delta(q^2_\perp)\sim \delta^2(\vec q_\perp)$
 if we take this part as a distribution of $q_\perp^2$. That is, this part
 may be factorized before the integration over $q_\perp$.  But, the part in the second line also give
 a divergent contribution to $A_N$ and it results in the discrepancy.
\par
To summarize: We have studied SSA in the lepton angular distribution of Drell-Yan processes. The asymmetry has been
studied with the diagram expansion at hadron level. Different results have been obtained. In this work,
we take a different approach. We calculate the transverse-spin dependent part of the differential cross-section
with suitable parton states. Because the spin is transverse, one has to take multi-parton states
for the purpose. Our result agrees with that in \cite{ZM}, but disagrees with those in \cite{DY1,DY2,DY3,AT}.
A possible reason for this has been discussed. It should be emphasized that the studied SSA is very interesting,
because it is at order of $\alpha_s^0$ and its prediction takes a simple form. Measuring such a SSA likely provides
the best way to access the twist-3 matrix element. Finally, we notice that there is also a corresponding SSA starting at order
of $\alpha_s^0$ in semi-inclusive DIS and its prediction takes a simple form. The work about this will appear elsewhere.

\par\vskip20pt
\noindent
{\bf Acknowledgments}
\par
This work is supported by National Nature
Science Foundation of P.R. China(No. 10975169, 11021092, 11275244).
\par\vskip30pt

\par

\end{document}